\documentclass[10pt]{IEEEtran}

\usepackage{amsmath,amssymb,amsthm,graphicx,graphics,times,multicol,citesort}

\newcommand{\cC}{\mathcal{C}}

\newcommand{\bp}{\mathbf{p}}

\newcommand{\br}{\mathbf{r}}

\newcommand{\be}{\mathbf{e}}
\newcommand{\bx}{\mathbf{x}}
\newcommand{\bn}{\mathbf{n}}

\newcommand{\bh}{\mathbf{h}}

\newcommand{\SNR}{\mathop{\mathsf{SNR}}}

\newcommand{\CN}{\mathop{\it CN}}
\newcommand{\argmax}{\mathop{\rm argmax}}

\newtheorem{Pro}{Proposition}

\begin{document}

\title{The Value of Staying Current when Beamforming}
\author{Yiyue Wu,  Andreas Achtzehn, Marina Petrova, Petri M$\ddot{\textrm{a}}$h$\ddot{\textrm{o}}$nen and Robert Calderbank}


\maketitle

\begin{abstract}
Beamforming is a widely used method of provisioning high quality wireless channels that leads to high data rates and simple decoding structures. It requires feedback of Channel State Information (CSI) from receiver to transmitter, and the accuracy of this information is limited by rate constraints on the feedback channel and by delay. It is important to understand how the performance gains associated with beamforming depend on the accuracy or currency of the Channel State Information.
This paper quantifies performance degradation caused by aging of CSI. It uses outage probability to measure the currency of CSI, and to discount the performance gains associated with ideal beamforming. Outage probability is a function of the beamforming algorithm and results are presented for Transmit Antenna Selection and other widely used methods. These results are translated into effective diversity orders for Multiple Input Single Output (MISO) and Multiuser Multiple Input Multiple Output (MIMO) systems.

\end{abstract}
\vspace{0.1cm}
\begin{IEEEkeywords}
Wireless Communication, Multiple Antennas, Beamforming, Channel State Information, Feedback, Outage Probability, Diversity Order
\end{IEEEkeywords}

\section{Introduction}


Spectral efficiency is critical to high-rate wireless communication and one way to achieve channel diversity is through the introduction of multiple antennas. The ability to vary transmission rate with the quality of the wireless channel requires that Channel State Information (CSI)  is available at the transmitter. When the forward channel from transmitter to receiver is different from the reverse channel, the CSI at the receiver needs to be reported to the transmitter via a feedback channel.


If latency is not a critical issue then system throughput is maximized by scheduling; users should request data only when their channel quality clears an appropriate threshold (see \cite{Viswanath-IT-02} for more details). When low latency is required, methods of engineering higher quality channels become important. There has been considerable recent interest in applying beamforming techniques, such as precoder selection, within 3G or WiMAX communication networks where channel diversity is a system resource. Note that provisioning higher quality channels can also be viewed in terms of reducing receiver complexity; when channel quality is good, the difference in performance between Maximum Likelihood (ML) decoding and suboptimal methods such as Zero Forcing (ZF) is very small \cite{Wu-ICASSP-10}.


The accuracy of CSI at the transmitter is limited by rate constraints on the feedback channel and by delay. The performance loss  associated with quantization of CSI at the receiver has been studied extensively (see \cite{MUkkavilli-IT-03,Love-IT-03,Santipach-IT-09,Wu-ICEAA-09}), but the loss associated with delay has received less attention. This paper analyzes the impact of feedback delay by using information theory to measure currency of CSI.

There are different sources of feedback delay. CSI is sent only periodically to minimize control overhead, and once received, the post-processing overhead in multi-user systems that is required for adaptation may also be significant.
Onggosanusi \textit{et al.} \cite{Onggosanusi-TC-01} measure the performance loss from feedback delay in terms of increased Bit Error Rate (BER) and show that there comes a point at which feedback hurts rather than helps. This is confirmed by Huang \textit{et al.} \cite{Huang-Globecomm-06} who measure the capacity of a system subject to feedback delay and conclude that it decreases at least exponentially with increasing delay.

In this paper we focus on outage probability rather than Bit Error Rate since interleaving and forward error correction cause changes in BER to trail changes in CSI. We would argue that it is also a good match to the systems objective of enabling ML decoding performance with ZF complexity.

First, we consider multiple-input and single-output (MISO) systems with different beamforming techniques. We start by considering Random Vector Quantization (RVQ) codebooks and by deriving an analytical expression for the outage probability of RVQ beamforming. This makes it possible to analyze the tradeoff between currency (feedback delay) and codebook size for different channel types. We also consider Transmit Antenna Selection (TAS) which is simple and widely used. To contrast the above findings we use outage probability estimations for perfect beamforming (PBF) with feedback delay explored by Annapureddy \textit{et al.} in \cite{Anna-Tcom-09}. In the following, we extend the results on perfect beamforming, RVQ beamforming and TAS to multiuser multiple input multiple output (MU-MIMO) systems.

The contributions of this paper are:

\begin{enumerate}
 \item {An outage probability analysis for systems with multiple transmit antennae and a single receiver antenna (MISO). We derive the outage probability for RVQ and TAS codebooks given SNR and the persistence properties of the channel.}
 \item {A numerical study showing that increasing the size of RVQ codebooks can reduce the loss from feedback delay. }
\item {A comparison of the performance of RVQ and TAS with conclusions on feasibility in different propagation environments. }
\item {The derivation of average outage probability for PBF, RVQ and TAS in the context of Multi-User MIMO (MU-MIMO) systems. We assume the receivers employ maximal ratio combining and we derive a baseline by extending results in \cite{Anna-Tcom-09,Ramya-ISIT-09} from MISO to MU-MIMO.}
\end{enumerate}

The rest of this paper is organized as follows. Section \ref{sysmod} presents the system and channel models for MISO, reformulating the selection problem for arbitrary codebooks to that of selecting the code that maximizes the parameter of a non-central chi-square distribution with two degrees of freedom. In Section \ref{SinUser} we study systems with delayed feedback and analyze the performance tradeoff between different beamforming schemes. The findings from Section \ref{SinUser} provide a foundation for the study of the MU-MIMO model described in Section \ref{MultiUser}. Conclusions are presented in Section \ref{conclusion}.

\section{System Overview}
\label{sysmod}
In a wireless communication system with $N_t$ transmit antennas and $N_r$ receive antennas, the received signal $\br$ is given by
\begin{equation}
\label{model0}
\small
\br=\tilde{\bh}  \bx^{\dag} + \bn
\end{equation}
where $\tilde{\bh}\in \mathbb{C}^{ N_t\times N_r}$ is the channel vector with the entry $\tilde{h}_{i,j}$ representing the complex channel gain between the $i^{th}$ transmit antenna and the $j^{th}$ receive antenna; $\bx$ represents the transmitted signal vector, and $\bn$ is the corresponding additive noise vector with each component following the complex Gaussian distribution, $\CN(0,1)$.
\subsection{Channel State Information Model}
Feedback delay results in CSI at the transmitter that is inaccurate, and the resulting mismatch between the true channel and the estimate causes errors in code selection. Note that it is possible to base feedback on long-range prediction of rapidly time-varying correlated fading channels (see \cite{Wu-ICEAA-09}) but this approach is outside the scope of our model. In our system model, we assume no forward prediction of CSI.

Jakes \cite{jakes} models the channel state evolution for time-selective narrowband channels as a Markov chain. The autocorrelation of the transfer function as a function of the time difference $\delta{t}$ is derived as
\begin{equation}
\label{ChannelModel}
\rho (\delta{t}) = J_0(2 \pi f_d\delta{t})
\end{equation}
where $J_0(\cdot)$ denotes the zero-order Bessel function of the first kind and $f_d$ is the maximum Doppler shift. Note that for small $\delta{t}$, the slope of the autocorrelation $\rho$ is strictly negative. We view $\rho$ as a measure of the persistence of the channel - the larger $\|\rho\|$, the stronger is the correlation between the latest channel measurement and the current channel state.

We model channel state persistence by a simple Markov chain as in \cite{marathe-asilomar-05,Nguyen-PIMRC-09}. We define the current channel state $\tilde{\bh}$ in relation to the previous channel estimation $\bh$ as
\begin{equation}
\label{model1}
\tilde{\bh} = \rho \bh + \sqrt{1-\rho^2}\be
\end{equation}
where $\be$ is the deviation from the estimate. The variables $\bh$ and $\be$ are independent and each of their components follows the standard complex Gaussian distribution, $\CN(0,1)$.

\section{Outage Analysis for MISO systems with Feedback Delay}\label{SinUser}

We first consider a MISO system with transmit beamforming. The system is composed of  $N_t$ transmit antennas and one receive antenna. Bandwidth occupancy is small in comparison to the coherence bandwidth. Let $\mathcal{C}=\{\bp_1, \cdots, \bp_N \}$ be an arbitrary (rank-one) beamforming codebook with $\bp_i \in \mathbb{C}^{N_t}$ and a uniform energy budget $\|\bp_i\|^2 = 1$. Thus the received signal for the beamforming system is written as
\begin{equation}
\label{model1}
\small
r= \langle\tilde{\bh}, \hat{\bp}\rangle x + n
\end{equation}
where $\langle\cdot,\cdot \rangle$ is the standard inner product and $\hat{\bp}$ is the selected beamformer.
In this section we estimate the average outage probability for different codebook schemes . In the following, we refer to the average outage probability as $P_{out}^C$ where $C$ is the codebook type referenced.

Outage is encountered when the mutual information between transmitted and received symbols is less than the system transmission rate. For the MISO system with the channel model in \eqref{model1}, the general instantaneous probability of system outage is given by
\begin{align}
\label{Outage1}
P_{\textrm{out}}(R,\epsilon ) & = \Pr\left[ \mathcal{I}(\br,\bx|\tilde{\bh}) < R  \right] \nonumber \\
& = \Pr\left[ \log_2(1 + \frac{\epsilon}{N_t} |\langle\tilde{\bh},\bp_i\rangle|^2) < R  \right] \nonumber \\
& = \Pr\left[ \left|\langle\tilde{\bh},\bp_i\rangle\right|^2 < \gamma_0  \right]
\end{align}
where $R$ is the transmission rate, $\epsilon$ is the signal to noise ratio ($\SNR$) and $\gamma_0 = \frac{2^R-1}{\epsilon/N_t}$. To incorporate feedback delay modeled in equation \eqref{ChannelModel}, we further write
\begin{align}
\label{outagegeneral}
P_{\textrm{out}}(R, \epsilon) & = \Pr\left[ \left|\rho\langle\bh ,\hat{\bp}\rangle + \sqrt{1-\rho^2}\langle\be,\hat{\bp}\rangle\right|^2 < \gamma_0  \right]\nonumber \\
& = \Pr\left[ \left|\sqrt{2\mu}\langle\bh ,\hat{\bp}\rangle + \sqrt{2}\langle\be,\hat{\bp}\rangle\right|^2 < \frac{2\gamma_0}{1-\rho^2}  \right] \nonumber\\
& = \Pr\left[ \left|\sqrt{2\mu\gamma} + \sqrt{2} z \right|^2 < \frac{2\gamma_0}{1-\rho^2}  \right]
\end{align}
where $\mu = \frac{\rho^2}{1-\rho^2}$, $\gamma=|\langle\bh ,\hat{\bp}\rangle|^2$ and $ z= \langle\be,\hat{\bp}\rangle e^{-i \angle \langle\bh,\hat{\bp}\rangle } $ is complex Gaussian distributed with zero mean and unit variance. So $\left| \sqrt{2\mu\gamma} + \sqrt{2} z \right|^2  $ is non-central chi-square distributed with two degrees of freedom and parameter $\sqrt{2\mu\gamma}$.

Equation \eqref{outagegeneral} provides us with the convenient means to study the average outage probability for different codebook schemes. The original selection problem for codes remains unchanged due to the duality in selecting the code that maximizes $\gamma$, but the left-hand side of equation \eqref{outagegeneral} lends itself to analytical methods as the average outage probability is tractable through the distribution of the instantaneous outage probability.

\subsection{MISO-PBF scenario}
In \cite{Anna-Tcom-09}, Annapureddy \textit{et al.} derive the outage probability for a beamforming system that is subject to feedback delay. They show that in the low $\SNR$ regime, power allocation in the direction of the CSI performs better, while uniform spatial power allocation (USPA) is beneficial in high $\SNR$ scenarios. The minimum outage probability for unconstrained codebook cardinality is
\begin{align}
& P^\textrm{PBF}_\textrm{out} (R, \epsilon, \rho)\nonumber \\
& \;\;\;= \frac{1}{(1+\mu)^{N_t-1}} \sum_{k=0}^{N_t-1} \left(\begin{array}{c} N_t-1\\k\end{array}\right) \frac{\mu^k}{k-1} \Gamma_{k+1}\left(\gamma_0\right)
\end{align}
where $\Gamma_{k}\left(x\right)$ is the lower incomplete gamma function.

\subsection{MISO-RVQ scenario}
RVQ codebooks are parameterized by their cardinality $N$ and consist of vectors that are isotropic (uniformly distributed over the sphere).
We define the tradeoff factor $\nu = \max_{\bp\in\cC}\frac{|\langle\bh,\bp\rangle|^2}{\|\bh\|^2}$ as the loss of adaptability in comparison to the perfect mutual-information maximizing code due to the limited size of the RVQ codebook. Complex Gaussian distribution of the channel vector and isotropy of the codebook yields the probability density function that appears in \cite{Yeung-TWC-09}

\begin{align}
\label{nupdf}
f_\nu = N(N_t-1)\left(1-(1-\nu)^{N_t-1}\right)^{N-1}(1-\nu)^{N_t-2}.
\end{align}

We now extend \eqref{outagegeneral} with the tradeoff factor and average over the probability density function, resulting in

\begin{align}
\label{bound_PBF}
& P^\textrm{RVQ}_\textrm{out} (R, \epsilon, \rho)\nonumber &= \sum_{k=0}^{N_t-1} A(k) \left(\begin{array}{c} N_t-1\\k\end{array}\right) \frac{\mu^k}{k-1} \Gamma_{k+1}\left(\gamma_0\right)
\end{align}
with
\begin{align}
A(k) = \int_{0}^{1}\frac{1}{(1+\mu\nu)^{N_t-1}} \nu^k f_\nu d\nu.
\end{align}

{\it Proof:} Given $\nu$, the outage probability is
\begin{align}
& \Pr(\left.\textrm{outage}\right|\nu) \nonumber \\
& = \Pr\left[ \left|\sqrt{2\mu\nu\gamma} + \sqrt{2}z \right|^2 < \frac{2\gamma_0}{1-\rho^2}  \right] \nonumber \\
& = \frac{1}{(1+\nu\mu)^{N_t-1}} \sum_{k=0}^{N_t-1} \left(\begin{array}{c} N_t-1\\k\end{array}\right) \frac{(\nu\mu)^k}{k-1} \Gamma_{k+1}\left(\frac{e^R -1}{\epsilon/N_t}\right).
\end{align}

Then, the overall outage probability for random vector quantization beamforming systems of $N$ random vectors is
\begin{align}
P_{\textrm{out}}^{\textrm{RVQ}} & = \int \Pr(\left.\textrm{outage}\right|\nu) f_\nu d\nu. \nonumber
\end{align}
$\square$\hfill

\begin{figure}[ht!]
\begin{center}
\resizebox{9cm}{!}{\includegraphics[scale=10]{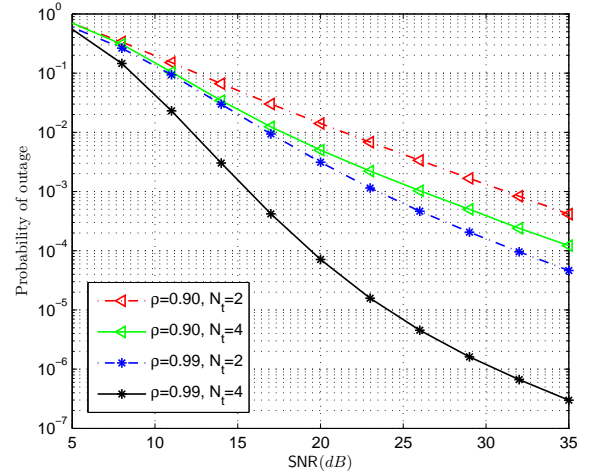}}
\caption{Outage Probability for $4\times1$ and $2\times1$ systems with a RVQ beamforming codebook of size eight and transmission rate at 2 bits/s/Hz.} \label{fig:OPRVQ}
\end{center}
\end{figure}

Fig. \ref{fig:OPRVQ} plots outage probability as a function of persistence; the RVQ codebook has cardinality 8, the transmission rate is 2 bits/s/Hz, and performance of a $4\times1$ MISO system is compared to that of a similar $2\times1$ MISO system. Note that loss of persistence results in curves with equal slopes, indicating a loss in MISO diversity order. 

\begin{figure}[ht!]
\begin{center}
 \resizebox{9cm}{!}{\includegraphics[scale=10]{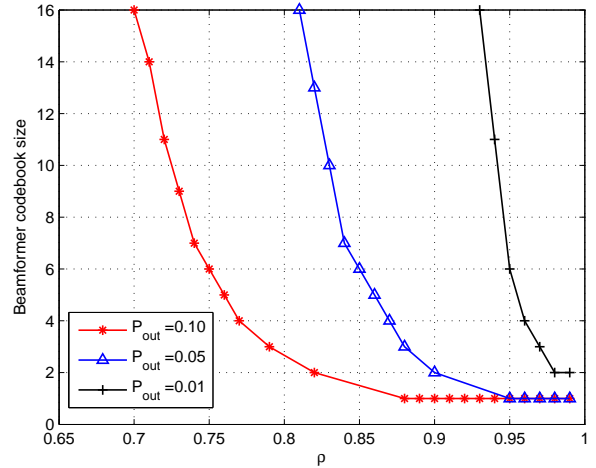}}
\caption{Minimum RVQ codebook size for given outage probability depending on channel persistence.}
 \label{fig:tradeoffNrho}
 \end{center}
\end{figure}

In the opposite direction, Fig. \ref{fig:tradeoffNrho} starts with a target outage probability and plots the required size of the RVQ codebook as a function of persistence. Note that outage probabilities of 1\% are representative of delay sensitive applications such as voice, whereas probabilities of 10\% or higher are representative of best effort data services. We see that codebook size grows exponentially with lower persistence and that the rate of growth depends on the target error probability. We expect to explore the value of adapting the choice of codebook in future work.

The diversity order is defined as
\begin{align}
D = - \lim_{\epsilon\rightarrow\infty}\frac{\log P_\textrm{out} (\epsilon)}{\log \epsilon}.
\end{align}

Using the fact that $\Gamma_{k}(x) \approx \frac{x^k}{k!}$, we approximate the outage probability in high $\SNR$ regime as
\begin{align}
P_\textrm{out}^\textrm{RVQ} (R, \epsilon, \rho)& \approx \gamma_0 \int_{0}^{1}\frac{1}{(1+\mu\nu)^{N_t-1}} (1 + \mu\gamma_0\nu)^{N_t-1} f_\nu d\nu \nonumber\\
& \approx \gamma_0 \int_{0}^{1}\frac{1}{(1+\mu\nu)^{N_t-1}} f_\nu d\nu
\end{align}
where $\gamma_0 = \frac{2^R-1}{\epsilon/N_t}$.

Therefore, the diversity order for RVQ beamforming in the presence of feedback delay ($0\leq\rho<1$) is
\begin{equation}
D^{\textrm{RVQ}}(0\leq\rho<1) = - \lim_{\epsilon\rightarrow\infty}\frac{\log \gamma_0}{\log \epsilon}=1.
\end{equation}

When there is no delay, the outage probability for RVQ beamforming is
\begin{align}
P_\textrm{out}^\textrm{RVQ}(R, \epsilon, \rho=1) = \int_{0}^{1}\Gamma_{N_t}\left(\frac{\gamma_0}{\nu}\right)  f_\nu d\nu.
\end{align}

So, the diversity order for RVQ beamforming with no feedback delay is given by
\begin{align}
D^{\textrm{RVQ}}(\rho=1) = - \lim_{\epsilon\rightarrow\infty}\frac{\log \gamma_0^{N_t}}{\log \epsilon}=N_t.
\end{align}
Hence, the diversity order for RVQ beamforming is
\begin{equation}
D^{\textrm{RVQ}} =\left\{\begin{array}{cc} 1, &  0\leq\rho<1,\\ \\ N_t, & \rho=1. \end{array}\right.
\end{equation}

\textbf{Remark:} In a open-loop wireless communication systems with $N_t$ transmit antennas and single receive antenna, space-time coding helps to achieve the diversity order of $N_t$ by scheduling streams across different transmit antennas. Our analysis is based on non-coded beamforming systems. With perfect feedback, it can achieve the diversity order $N_t$ and the diversity deteriorates to 1 as feedback becomes delayed. Note that, in the case of delayed feedback, space-time coding can also helps to obtain full diversity.

\begin{figure}[ht!]
\begin{center}
\resizebox{9cm}{!}{\includegraphics[scale=10]{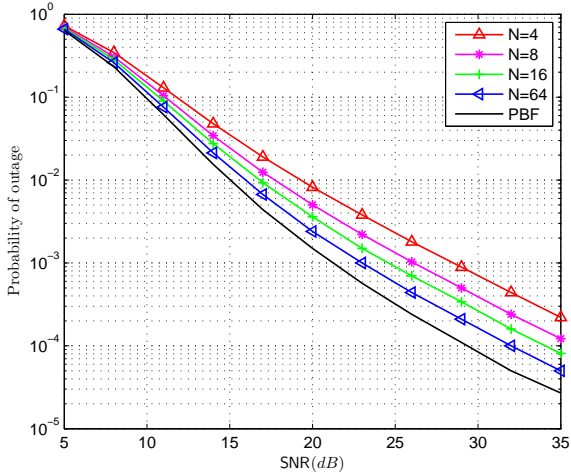}}
\caption{Outage Probability for $4\times1$ systems with a RVQ beamforming codebook of varying size, transmission rate at 2 bits/s/Hz and $\rho=0.90$; in the legend, `PBF' denotes perfect beamforming.} \label{fig:OPRVQ_1}
\end{center}
\end{figure}

In Figure \ref{fig:OPRVQ_1} we study the possibility of mitigating the degradation effect by increasing the codebook size. Again, we assume a $4\times1$ MISO setup and plot the outage probability for different codebook sizes at a fixed channel persistence of $\rho = 0.9$. Observe that larger codebooks converge towards the perfect beamforming case, but cannot mitigate the loss in diversity order.

\subsection{MISO-TAS scenario}
In pure TAS scenarios, the transmitter-side antenna is selected for which the channel gain is maximal, hence  $$ \gamma_{\textrm{TAS}} = \max_{i}|h_i|^2.$$
$\gamma_{\textrm{TAS}}$ is the largest member of a set of exponentially distributed channel gain values $h_i$ with probability density function

\begin{align}
f_{\gamma_{\textrm{TAS}}} = N_t\left(1-e^{-x}\right)^{N_t-1} e^{-x}.
\end{align}

Based on the distribution of $\gamma_{\textrm{TAS}}$,  we now derive the outage probability for TAS scenarios.
\begin{Pro}[MISO-TAS outage probability] \label{p1}
Consider a $N_t \times 1$ wireless fading channel employing transmit antenna selection and transmitting the signals at a date rate of $R$ bits/s/Hz with $\SNR = \epsilon$, then the outage probability in the presence of feedback delay is given by
\begin{align}
\label{OPTAS}
 \mbox{\fontsize{8}{10}\selectfont $\displaystyle P_{\textrm{out}}^{\textrm{TAS}}(R, \epsilon, \rho)  = N_t \sum_{k=0}^{N_t-1} \left(\begin{array}{c}N_t-1 \\ k\end{array}\right) \frac{(-1)^k}{k+1} \left(1-e^{-\frac{k+1}{k+1+u}\frac{2\gamma_0}{1-\rho^2}}\right)$}
\end{align}
where $\gamma_0 = \frac{2^R-1}{\epsilon/N_t}$.
\end{Pro}

{\it Proof:} We recall from equation \eqref{outagegeneral} that $\left| \lambda + \sqrt{2} z \right|^2  $ is non-central chi-square distributed. For a given $\gamma_{\textrm{TAS}}$ we can therefore derive equality

\begin{align}
\Pr(\left.\textrm{outage}\right|\lambda) &=\Pr\left[ \left|\sqrt{2\mu\gamma} + \sqrt{2} z \right|^2 < \frac{2\gamma_0}{1-\rho^2}  \right] \nonumber \\
& = F_{(\textrm{nc}-\mathcal{X}^2, 2, 2\mu\gamma)}\left(2\beta\right)
\end{align}
where $\beta=\frac{\gamma_0}{1-\rho^2}$ and $F_{(\textrm{nc}-\mathcal{X}^2,\; n,\; m)}(\cdot)$ is the cumulative probability function for the non-central chi-square distribution with $n$ degrees of freedom and parameter $m$.

So the overall outage probability is
\begin{align}
P_{\textrm{out}}^{\textrm{TAS}} = \int \Pr(\left.\textrm{outage}\right|\lambda)f_\lambda(\lambda)d\lambda.
\end{align}

Using the fact that
\begin{align}
\label{nc-chi-cdf}
F_{(\textrm{nc}-\mathcal{X}^2, 2d, 2\delta)\left(2\beta\right)} = \sum_{k=0}^{\infty}\frac{(\delta)^k e^{-\delta}}{k!} F_{(\mathcal{X}^2,2d+2k)}\left(2\beta\right)
\end{align}
where $F_{(\mathcal{X}^2,2+2k)}(\cdot)$ is the cumulative probability function of chi-square distribution with degree $2 + 2k$ and the fact that
\begin{align}
\label{chi-cdf}
F_{(\mathcal{X}^2,2d+2k)}\left(2\beta\right) = \int_0^\beta\frac{x^{k+d-1} e^{-x}}{(k+d-1)!}dx,
\end{align}

we can rewrite the outage probability as
\begin{align}
& P_{\textrm{out}}^{\textrm{TAS}} \nonumber \\
& = \mbox{\fontsize{8}{10}\selectfont $\displaystyle\sum_{k=0}^{\infty} \frac{\mu^k}{k!} \int_0^\beta\frac{y^ke^{-y}}{k!}dy \int_0^{\infty} N_t\left(1-e^{-x}\right)^{N_t-1} x^k e^{-(1+\mu) x} dx$} \nonumber \\
& = \mbox{\fontsize{8}{10}\selectfont $\displaystyle N_t\sum_{k=0}^{\infty} \frac{\mu^k}{k!} \int_0^\beta\frac{y^ke^{-y}}{k!}dy  \sum_{j=1}^{N_t-1}  \frac{\left(\begin{array}{c}N_t-1 \\ j\end{array}\right) (-1)^jk!}{(j+1+\mu)^{k+1}}$} \nonumber \\
& = \mbox{\fontsize{8}{10}\selectfont $\displaystyle N_t \sum_{j=1}^{N_t-1} \left(\begin{array}{c}N_t-1 \\ j\end{array}\right) \frac{(-1)^j}{j+1+\mu} \int_0^\beta e^{-y} \sum_{k=0}^{\infty}   \frac{(\frac{\mu y}{j+1+\mu})^k }{k!}dy $}   \nonumber\\
& =  N_t \sum_{j=1}^{N_t-1} \left(\begin{array}{c}N_t-1 \\ j\end{array}\right) \frac{(-1)^j}{j+1+\mu} \int_0^\beta e^{-\frac{(j+1) y}{j+1+\mu}}dy \;   \nonumber \\
& = N_t \sum_{k=0}^{N_t-1} \left(\begin{array}{c}N_t-1 \\ k\end{array}\right) \frac{(-1)^k}{k+1} \left(1-e^{-\frac{k+1}{k+1+u}\frac{2\gamma_0}{1-\rho^2}}\right).
\end{align}
$\square$\hfill

\begin{figure}[ht!]
\begin{center}
\resizebox{9cm}{!}{\includegraphics[scale=10]{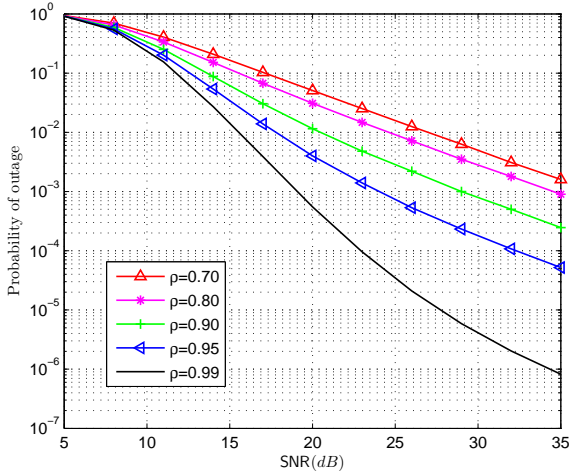}}
\caption{Outage Probability for $4\times1$ systems with transmit antenna selection and transmission rate 2 bits/s/Hz.} \label{fig:OPTAS}
\end{center}
\end{figure}

Fig. \ref{fig:OPTAS} plots outage probability for TAS beamforming as a function of persistence.
The pattern is similar to that for RVQ codebooks; loss of persistence results in curves with equal slopes, indicating a loss in MISO diversity order

In the presence of feedback delay ($0 < \rho < 1$), the diversity order for TAS is
\begin{align}
\mbox{\fontsize{8}{10}\selectfont $\displaystyle \lim_{\epsilon\rightarrow\infty} -\frac{\log\left(\gamma_0 N_t\displaystyle \sum_{k=0}^{N_t-1} \left(\begin{array}{c}N_t-1 \\ k\end{array}\right) \frac{2(-1)^k}{(k+1+u)(1-\rho^2)}\right)}{\log\epsilon} = 1$}.
\end{align}
Hence, the diversity order is given by
\begin{equation}
D^{\textrm{TAS}} =\left\{\begin{array}{cc} 1, &  0\leq\rho<1,\\ \\ N_t, & \rho=1. \end{array}\right.
\end{equation}

\subsection{Comparison}
\begin{figure}[h!]
\begin{center}
\resizebox{9cm}{!}{\includegraphics[scale=10]{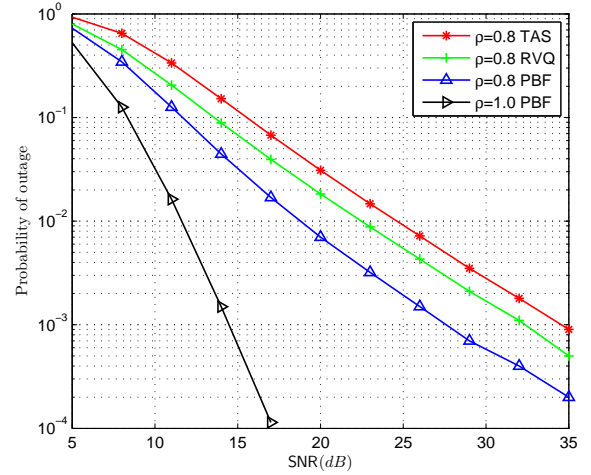}}
\caption{Outage Probability for $4\times1$ systems with different beamforming schemes; transmission rate is 2 bits/s/Hz and the RVQ beamforming codebook has eight beamforming vectors.} \label{fig:OPCompare}
\end{center}
\end{figure}

Fig. \ref{fig:OPCompare} plots outage probability in a low-persistence channel ($\rho = 0.8$) as a function of SNR. As expected, Perfect Beamforming (PBF) is superior to RVQ beamforming and Transmit Antenna Selection (TAS). However the gap between PBF and RVQ will narrow as the size of the codebook increases. When the persistence $\rho<1$, all schemes experience a loss in diversity order.


\section{Outage analysis for multiuser MIMO systems with Feedback Delay}
\label{MultiUser}
Now we consider a $N_u$-user system with a base station employing $N_t$ transmit antennas and each user equipped with $N_r$ receive antennas. We first derive the outage probabilities for multiuser MIMO systems with transmit antenna selection. Then, using the duality between perfect beamforming at the transmitter and maximal ratio combining at the receiver, we will derive the outage probabilities for multiuser MISO systems with perfect beamforming and RVQ beamforming.
\subsection{MU-MIMO TAS Scenario}
The outage probability for transmit antenna selection with multiuser diversity in the case of no-delay feedback is given in \cite{Zhang-SPL-09}. We extend the analysis to the case of delayed feedback. During each coherent interval, the scheduler selects the $i^{th}$ transmit antenna and the $k^{th}$ user by
\begin{align}
\{\hat{i}, \hat{k}\} = \argmax_{i,k} \|\bh_{i}^{(k)}\|^2
\end{align}
where $\bh_{i}^{(k)}=\left(h_{i,1}^{(k)},\cdots, h_{i,N_r}^{(k)}\right)$ represents the channel vector between the $i^{th}$ transmit antenna and the $k^{th}$ user. Given Maximal Ratio Combining (MRC) at the $\hat{k}^{th}$ user, the received signal $\br^{(\hat{k})}$ is given by
\begin{equation}
\br^{(\hat{k})} = \|\bh_{\hat{i}}^{(\hat{k})}\|^2 x + \sum_{j=1}^{N_r}\langle\bh_{\hat{i},j}^{(\hat{k})},\bn_{\hat{i},j}^{(\hat{k})}\rangle
\end{equation}
where $\sum_{j=1}^{N_r}\langle\bh_{\hat{i},j}^{(\hat{k})},\bn_{\hat{i},j}^{(\hat{k})}\rangle$ is the combined noise with zero mean and variance $N_r$. The cumulative probability function of $\eta=\|\bh_{\hat{i}}^{(\hat{k})}\|^2$ is given by
\begin{align}
F_{\eta}(x) = \left(1-e^{-x}\sum_{n=0}^{Nr-1}\frac{x^n}{n!}\right)^Z
\end{align}
where $Z = N_u N_t$ and its probability density function is given by
\begin{align}
\label{etapdf}
f_{\eta}(x) = \frac{Z}{(N_r-1)!} x^{N_r-1} e^{-x} \left(1-e^{-x}\sum_{n=0}^{Nr-1}\frac{x^n}{n!}\right)^{Z-1}.
\end{align}

The outage probability in the presence of feedback delay is
\begin{align}
\label{OutageMUTAS1}
P_{\textrm{out}}(R,\epsilon ) & = \Pr\left[ \|\tilde{\bh}_{\hat{i}}^{(\hat{k})}\|^2 < \frac{2^R-1}{\epsilon/N_t}  \right] \nonumber \\
& = \Pr\left[ \sum_{j=1}^{N_r}\left|\sqrt{2\mu|\bh_{\hat{i},j}^{(\hat{k})}|^2} + \sqrt{2}z_j \right|^2 < \frac{2\gamma_0}{1-\rho^2}  \right]. \nonumber\\
\end{align}
where $\tilde{\bh}_{\hat{i}}^{(\hat{k})}$ is defined in \eqref{ChannelModel} and $z_j =\be^{(\hat{k})}_{\hat{i},j} e^{-i \angle \bh_{\hat{i},j}^{(\hat{k})}}$ is zero mean complex Gaussian with unit variance. Note that $\sum_{j=1}^{N_r}\left|\sqrt{2\mu|\bh_{\hat{i},j}^{(\hat{k})}|^2} + \sqrt{2}z_j \right|^2$ is non-central chi-square distributed with $2N_r$ degrees of freedom and parameter $\sqrt{2\mu\eta}$.

Using the distribution in \eqref{nc-chi-cdf} and \eqref{chi-cdf}, we write the outage probability for a given $\eta$ as
\begin{align}
\Pr(\left.\textrm{outage}\right|\eta) & = F_{(\textrm{nc}-\mathcal{X}^2, 2N_r, 2\mu\gamma)}\left(2\beta\right) \nonumber \\
& = \sum_{k=0}^{\infty}\frac{(\mu\gamma)^k e^{-\mu\gamma}}{k!} F_{(\mathcal{X}^2,2N_r+2k)}\left(2\beta\right) \nonumber \\
& = \sum_{k=0}^{\infty}\frac{(\mu\gamma)^k e^{-\mu\gamma}}{k!} \int_0^\beta\frac{x^{k+N_r-1} e^{-x}}{(k+N_r-1)!}dx.
\end{align}

Then, the overall outage probability for multiuser MIMO transmit antenna selection with MRC at the receiver is
\begin{align}
P_{\textrm{out}}^{\textrm{MUTAS}} = \int \Pr(\left.\textrm{outage}\right|\eta)f_\eta(\eta)d\eta.
\end{align}

{\it Lemma 1} \cite{Knuth68}: Let $m$, $n$, $k$ be positive integers and $m\geq n$, then the following equation holds:
\begin{align}
\label{Lemma1}
\left(\begin{array}{c} m+k\\n+k \end{array}\right) = \sum_{i=0}^{\min\{k,m-n\}} \left(\begin{array}{c}k\\i\end{array}\right) \left(\begin{array}{c}m\\i+n\end{array}\right).
\end{align}
$\square$\hfill



Now, we derive the outage probability for the multiuser MIMO systems with transmit antenna selection and MRC at the receiver.
\begin{Pro}[MU-MIMO TAS outage probability] \label{p2}
Consider a $N_u$-user wireless communication with the base station employing $N_t$ transmit antennas and each user employing $N_r$ receiver antennas employing transmit antenna selection and maximal ratio combining with the transmission date rate of $R$ bits/s/Hz and $\SNR = \epsilon$, the outage probability in the presence of feedback delay is given by
\begin{align}
\label{OPMUTAS}
& P_{\textrm{out}}^{\textrm{MUTAS}}(R, \epsilon, \rho) \nonumber \\
& \mbox{\fontsize{7}{8}\selectfont $\displaystyle  = \frac{N_u N_t}{(N_r-1)!}  \sum_{k=0}^{N_u N_t-1} \left(\begin{array}{c}N_u N_t-1 \\ k\end{array}\right) (-1)^{k} \sum_{m=0}^{k(N_r-1)}  \frac{m!\; a_m(N_r,k)}{(1+k+\mu)^m} $} \nonumber \\
& \mbox{\fontsize{7}{8}\selectfont $\displaystyle  \sum_{n=0}^m \frac{\mu^n (N_r+n-1)!}{n!(1+k)^{N_r+n}} \left(\begin{array}{c} N_r+m-1\\N_r+n-1 \end{array}\right)\Gamma_{N_r + n}\left(\frac{1+k}{1+k+\mu}\beta\right).$}
\end{align}
where $\beta = \frac{\gamma_0}{1-\rho^2}$ and $a_m(N_r,k)$ is defined in
\begin{align}
\left(\sum_{l=0}^{N_r-1}\frac{x^l}{l!}\right)^k = \sum_{m=0}^{k(N_r-1)}a_m(N_r,k)x^m.
\end{align}
\end{Pro}

{\it Proof:} the proof is provided in appendix which uses the result in Lemma 1.$\square$\hfill

The diversity order of multiuser MIMO systems with transmit antenna selection and receive maximal ratio combining can be derived as
\begin{equation}
D^{\textrm{MUTAS}} =\left\{\begin{array}{cc} N_r, &  0\leq\rho<1, \\ \\ N_uN_tN_r, & \rho=1. \end{array}\right.
\end{equation}

\textbf{Remark:} Proposition 2 is a special case of proposition 3 by setting $N_u=N_r=1$.

\begin{figure}[h!]
\begin{center}
\resizebox{9cm}{!}{\includegraphics[scale=10]{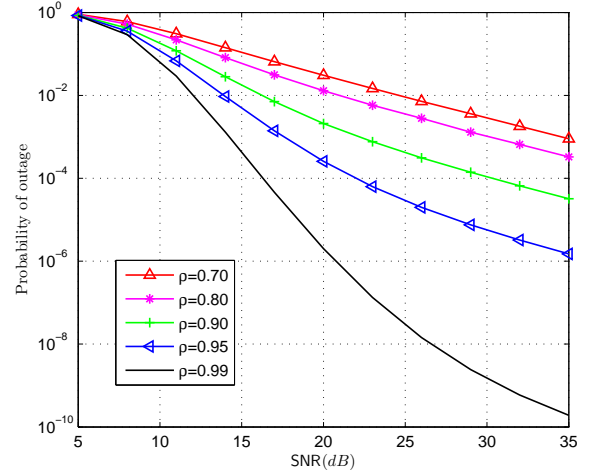}}
\caption{Outage Probability for 2-user systems with 4 transmit antennas at the base station and 2 receive antenna at each user; the transmission rate is 2 bits/s/Hz.} \label{fig:OPMUTAS}
\end{center}
\end{figure}

\begin{figure}[h!]
\begin{center}
\resizebox{9cm}{!}{\includegraphics[scale=10]{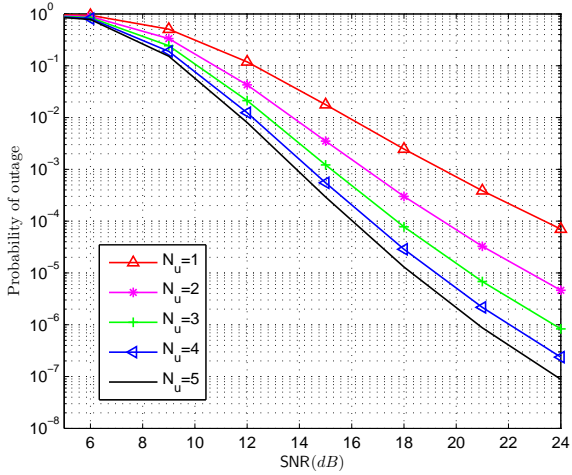}}
\caption{Outage Probability for multiuser systems with 4 transmit antennas at the base station and 2 receive antenna at each user; the transmission rate is 2 bits/s/Hz.} \label{fig:OPMUTAS_DiffUsers}
\end{center}
\end{figure}

It is shown in Fig. \ref{fig:OPMUTAS} that feedback delay has a great impact on transmit antenna selection in multiuser systems, where it considers 2-user wireless communication system with four transmit antennas at the base station and two receive antenna for each user. Fig. \ref{fig:OPMUTAS_DiffUsers} illustrates the outage comparison for systems with different users where their outage probabilities share the same diversity order.

\subsection{Outage Probabilities for multiuser MISO systems with perfect beamforming and RVQ beamforming}
In a $N_u$-user system with the base station employing $N_t$ transmit antennas and each user equipped with single receive antenna, the base station select the $\hat{k}^{th}$ user via the following algorithm:
\begin{align}
\hat{k} = \argmax_{k} \|\bh^{(k)}\|^2.
\end{align}
Then $\tau=\|\bh^{(\hat{k})}\|^2$ has the following cumulative density function
\begin{align}
F_{\eta}(x) = \left(1-e^{-x}\sum_{n=0}^{N_t-1}\frac{x^n}{n!}\right)^{N_u}
\end{align}
and its probability density function is given by
\begin{align}
\label{etapdf}
f_{\eta}(x) = \frac{N_u}{(N_t-1)!} x^{N_t-1} e^{-x} \left(1-e^{-x}\sum_{n=0}^{N_t-1}\frac{x^n}{n!}\right)^{N_u-1}.
\end{align}

Following the similar calculation for proposition 3, we can derive the outage probability for multiuser MISO systems with perfect beamforming as
\begin{align}
\label{OPMUPBF}
& P_{\textrm{out}}^{\textrm{MUPBF}}(R, \epsilon, \rho) \nonumber \\
& = \mbox{\fontsize{8}{10}\selectfont $\displaystyle \frac{N_u }{(N_t-1)!}  \sum_{k=0}^{N_u -1}  (-1)^{k}\left(\begin{array}{c}N_u -1 \\ k\end{array}\right) \sum_{m=0}^{k(N_t-1)}  \frac{m!\; a_m(N_t,k)}{(1+k+\mu)^m}$} \nonumber \\
& \mbox{\fontsize{8}{10}\selectfont $\displaystyle \sum_{n=0}^m \frac{\mu^n (N_t+n-1)!}{n!(1+k)^{N_t+n}} \left(\begin{array}{c} N_t+m-1\\N_t+n-1 \end{array}\right)\Gamma_{N_t + n}\left(\frac{(1+k)\beta}{1+k+\mu}\right)$}.
\end{align}
Noticing the duality between perfect beamforming for a $M\times1$ system and maximal ratio combining for a $1\times M$ system, the result in equation \eqref{OPMUPBF} can also be easily derived by switching $N_r$ and $N_t$ in equation \eqref{OPMUTAS}.

Then, the outage probability for  multiuser MISO systems with RVQ beamforming can be easily derived from equation \eqref{OPMUPBF} as
\begin{align}
\label{OPMURVQ}
& P_{\textrm{out}}^{\textrm{MURVQ}}(R, \epsilon, \rho) \nonumber \\
& = \mbox{\fontsize{8}{10}\selectfont $\displaystyle \frac{N_u }{(N_t-1)!}  \sum_{k=0}^{N_u -1}  (-1)^{k}\left(\begin{array}{c}N_u -1 \\ k\end{array}\right) \sum_{m=0}^{k(N_t-1)}m!\; a_m(N_t,k)  $} \nonumber \\
& \mbox{\fontsize{8}{10}\selectfont $\displaystyle\;\;  \sum_{n=0}^m\frac{ \mu^n(N_t+n-1)!}{n!(1+k)^{N_t+n}} \left(\begin{array}{c} N_t+m-1\\N_t+n-1 \end{array}\right)
\int_0^1 \frac{\nu^n\Gamma_{N_t + n}\left(\frac{(1+k)\beta}{1+k+\nu\mu}\right)}{(1+k+\nu\mu)^m}f_\nu d\nu$}.
\end{align}
where $f_\nu$ is defined in equation \eqref{nupdf}.

The diversity order for multiuser MISO systems with perfect beamforming and RVQ beamforming can be given by
\begin{equation}
D^{\textrm{MUBF}} =\left\{\begin{array}{cc} N_t, &  0\leq\rho<1, \\ \\ N_uN_t, & \rho=1. \end{array}\right.
\end{equation}

\section{Conclusion}
\label{conclusion}


In this paper we extended the current view on the effects of feedback imperfections with a framework for studying the effects of delay in the feedback process. We subsequently applied our findings to well-known beamforming techniques such as Perfect Beamforming, Random Vector Quantization and Transmit Antenna Selection, once more proving that delay is crucial to system performance.

Our new framework is based on the derivation of outage probabilities. We argued that this metric is most adequate to model the performance perception at different levels of the communication entities. Additionally, we studied the observable loss in diversity order and we found that diversity order decreases fast if delay is introduced, marginalizing the gains of complex antenna setups.

We evaluated through simulation the statistics of a single-user MISO system and a multi-user MIMO system. Here, we could show that an increase in the codebook size is to some extend capable of mitigating the performance degradation experienced due to the delay. This interesting finding allowed us to conduct a trade-off analysis between allowable delay and codebook volume.

Future work will further explore the value of adapting the codebook and/or its parameters to the delay environment. We expect to provide a more robust system architectural theory and mechanisms for environments with inherent delay.

\section*{Appendix}
\begin{center}
Proof of proposition 3
\end{center}

{\it Proof:} the outage probability for multiuser MIMO systems with transmit antenna selection and maximal ratio combining at the receiver is
\begin{align}
& P_{\textrm{out}}^{\textrm{MUTAS}} \nonumber \\
& = \int \Pr(\left.\textrm{outage}\right|\eta)f_\eta(\eta)d\eta. \nonumber\\
 & = \int_{\eta=0}^{\infty}\;\sum_{k=0}^{\infty}\frac{(\mu\eta)^k e^{-\mu\eta}}{k!} \int_0^\beta\frac{x^{k+N_r-1} e^{-x}}{(k+N_r-1)!}dx \; f_\eta (\eta) d \eta
\end{align}
Applying the probability density distribution $f_\eta$ in equation \eqref{etapdf}, we can further write the outage probability as
\begin{align}
\label{MUTASproof1}
& P_{\textrm{out}}^{\textrm{MUTAS}} \nonumber \\
 & = \mbox{\fontsize{8}{10}\selectfont $\displaystyle\sum_{j=0}^{\infty}\frac{\mu^j}{j!}\int_0^\beta \frac{x^{N_r+j-1} e^{-x}}{(N_r+j-1)!}dx \frac{Z}{(N_r-1)!} \sum_{k=0}^{Z-1} \left(\begin{array}{c} Z-1\\ k \end{array}\right)$}\nonumber \\
 &\mbox{\fontsize{8}{10}\selectfont $\displaystyle\;\; (-1)^k\sum_{m=0}^{k(N_r-1)} a_{m}(N_r,k) \int_0^{\infty} \eta^{N_r+m+j-1} e^{-(1+k+\mu)} d\eta$} \nonumber \\ \vspace{3cm}
 & = \mbox{\fontsize{8}{10}\selectfont $\displaystyle \sum_{j=0}^{\infty}\frac{\mu^j}{j!}\int_0^\beta \frac{x^{N_r+j-1} e^{-x}}{(N_r+j-1)!}dx \frac{Z}{(N_r-1)!} \sum_{k=0}^{Z-1} \left(\begin{array}{c} Z-1\\ k \end{array}\right)$} \nonumber \\
 &\;\; (-1)^k \sum_{m=0}^{k(N_r-1)} a_{m}(N_r,k) \frac{(N_r+m+j-1)!}{(1+k+\mu)^{N_r+m+j}} \nonumber \\
 & = \mbox{\fontsize{8}{10}\selectfont $\displaystyle \int_0^\beta x^{N_r-1} e^{-x}dx \frac{Z}{(N_r-1)!} \sum_{k=0}^{Z-1}(-1)^k \left(\begin{array}{c} Z-1\\ k \end{array}\right)$} \nonumber \\
 & \mbox{\fontsize{8}{10}\selectfont $\displaystyle \sum_{m=0}^{k(N_r-1)}\frac{ a_{m}(N_r,k)}{(1+k+\mu)^{N_r+m}} \sum_{j=0}^{\infty}\frac{\left(\frac{\mu x}{1+k+\mu}\right)^j}{j!} \frac{(N_r+m+j-1)!}{(N_r+j-1)!}.$}
\end{align}

To further simplify the equation in \eqref{MUTASproof1}, now we use Lemma 1 to establish the following equation:
\begin{align}
& \sum_{j=0}^{\infty}\frac{\left(\frac{\mu x}{1+k+\mu}\right)^j}{j!} \frac{(N_r+m+j-1)!}{(N_r+j-1)!} \nonumber\\
& = \mbox{\fontsize{8}{10}\selectfont $\displaystyle m! \sum_{j=0}^{\infty}\frac{\left(\frac{\mu x}{1+k+\mu}\right)^j}{j!} \left(\begin{array}{c} N_r+m+j-1 \\ N_r+j-1 \end{array}\right)$} \nonumber \\
& = \mbox{\fontsize{8}{10}\selectfont $\displaystyle m! \sum_{n=0}^{m}\sum_{j-n=0}^{\infty}\frac{\left(\frac{\mu x}{1+k+\mu}\right)^j}{j!} \left(\begin{array}{c} j \\ n \end{array}\right) \left(\begin{array}{c} N_r+m-1 \\ N_r+n-1 \end{array}\right)$} \nonumber \\
& = \mbox{\fontsize{8}{10}\selectfont $\displaystyle m! \sum_{n=0}^{m}\left(\frac{\mu x}{1+k+\mu}\right)^n \sum_{j=0}^{\infty}\frac{\left(\frac{\mu x}{1+k+\mu}\right)^j}{j!n!} \left(\begin{array}{c} N_r+m-1 \\ N_r+n-1 \end{array}\right)$} \nonumber \\
& = \mbox{\fontsize{8}{10}\selectfont $\displaystyle m! e^{\frac{\mu x}{1+k+\mu}} \sum_{n=0}^{m} \left(\frac{\mu x}{1+k+\mu}\right)^n \frac{1}{n!} \left(\begin{array}{c} N_r+m-1 \\ N_r+n-1 \end{array}\right)$}.
\end{align}

So, we can further derive the equation in \eqref{MUTASproof1} as
\begin{align}
 & P_{\textrm{out}}^{\textrm{MUTAS}} \nonumber \\
 & = \mbox{\fontsize{8}{10}\selectfont $\displaystyle \frac{Z}{(N_r-1)!} \sum_{k=0}^{Z-1}(-1)^k \left(\begin{array}{c} Z-1\\ k \end{array}\right) \sum_{m=0}^{k(N_r-1)}\frac{ a_{m}(N_r,k)m!}{(1+k+\mu)^{N_r+m}}$}  \nonumber \\
& \mbox{\fontsize{8}{10}\selectfont $\displaystyle \sum_{n=0}^{m}  \frac{\left(\begin{array}{c} N_r+m-1 \\ N_r+n-1 \end{array}\right)}{n!}  \int_0^\beta x^{N_r+i-1} \left(\frac{\mu x}{1+k+\mu}\right)^n e^{-\frac{(1+k) x}{1+k+\mu}} dx $} \nonumber\\
& = \mbox{\fontsize{8}{10}\selectfont $\displaystyle \frac{N_u N_t}{(N_r-1)!}  \sum_{k=0}^{N_u N_t-1} (-1)^{k}\left(\begin{array}{c}N_u N_t-1 \\ k\end{array}\right) \sum_{m=0}^{k(N_r-1)}  \frac{m!\; a_m(N_r,k)}{(1+k+\mu)^m}$} \nonumber \\
& \mbox{\fontsize{8}{10}\selectfont $\displaystyle \sum_{n=0}^m \frac{\mu^n (N_r+n-1)!}{n!(1+k)^{N_r+n}} \left(\begin{array}{c} N_r+m-1\\N_r+n-1 \end{array}\right)\Gamma_{N_r + n}\left(\frac{(1+k)\beta}{1+k+\mu}\right).$}
\end{align}
$\square$\hfill

\bibliographystyle{IEEEtran}
\bibliography{FD}

\end{document}